\documentclass[manuscript=letter, twocolumn]{revtex4-1} 

\usepackage{graphicx}
\usepackage[export]{adjustbox}
\usepackage{amsmath}
\usepackage{subfig}
\usepackage[justification=justified]{caption}
\usepackage{verbatim}
\usepackage{color}
\usepackage[section]{placeins}

\newcommand{\ket}[1]{| #1 \rangle}

\newcommand{\rr}{\mathbf{r}}

\newcommand{\q}{\mathbf{q}}

\newcommand{\m}{\mathrm}
\newcommand{\mb}{\mathbf}

\newcommand{\gowo}{G$_0$W$_0$~}

\usepackage[abs]{overpic}

\begin{document} 
\title{Quasiparticle band structure engineering in van der Waals heterostructures \emph{via} dielectric screening}

\author{Kirsten T. Winther}
\affiliation{Center for Atomic-scale Materials Design, Department of
Physics, Technical University of Denmark, DK - 2800 Kgs. Lyngby, Denmark}
\affiliation{Center for Nanostructured Graphene, Technical University of Denmark, DK - 2800 Kgs. Lyngby, Denmark}

\author{Kristian S. Thygesen}
\email{thygesen@fysik.dtu.dk}
\affiliation{Center for Atomic-scale Materials Design, Department of Physics, Technical University of Denmark, DK - 2800 Kgs. Lyngby, Denmark}
\affiliation{Center for Nanostructured Graphene, Technical University of Denmark, DK - 2800 Kgs. Lyngby, Denmark}


\begin{abstract}
The idea of combining different two-dimensional (2D) crystals in van der Waals heterostructures (vdWHs) has led to a new paradigm for band structure engineering with atomic precision. Due to the weak interlayer couplings, the band structures of the individual 2D crystals are largely preserved upon formation of the heterostructure. However, regardless of the details of the interlayer hybridisation, the size of the 2D crystal band gaps are always reduced due to the enhanced dielectric screening provided by the surrounding layers. The effect can be on the order of electron volts, but its precise magnitude is non-trivial to predict because of the non-local nature of the screening in quasi-2D materials, and it is not captured by effective single-particle methods such as density functional theory. Here we present an efficient and general method for calculating the band gap renormalization of a 2D material embedded in an arbitrary vdWH. The method evaluates the change in the GW self-energy of the 2D material from the change in the screened Coulomb interaction. The latter is obtained using the quantum-electrostatic heterostructure (QEH) model. We benchmark the G$\Delta$W method against full first-principles GW calculations and use it to unravel the importance of screening-induced band structure renormalisation in various vdWHs. A main result is the observation that the size of the band gap reduction of a given 2D material when inserted into a heterostructure scales inversely with the polarisability of the 2D material. Our work demonstrates that dielectric engineering \emph{via} van der Waals heterostructuring represents a promising strategy for tailoring the band structure of 2D materials. 
\end{abstract}

\maketitle


Atomically thin two-dimensional (2D) materials are being intensely scrutinized due to their unique and tuneable properties which make them candidates for high-performance applications in electronics, opto-electronics, and energy conversion\cite{ferrari2015science}. Following the discovery of graphene, a variety of other 2D materials have been isolated and characterized with the most well known examples being the insulator hexagonal boron-nitride (hBN), the elemental semiconductor phosphorene, and the semiconducting Mo- and W-based transition metal dichalcogenides (TMDs)\cite{Wang2012,castellanos2014isolation}. However, today the family of 2D materials counts numerous other members including further elemental crystals (Xenes), a variety of semiconducting and metallic TMDs, group III-V compounds, as well as transition metal carbides and -nitrides (MXenes), just to mention some\cite{bhimanapati2015recent}.

Even larger diversity can be achieved by stacking different 2D crystals into van der Waals heterostructures (vdWH)\cite{Geim2013}. The concept has proved itself extremely versatile and the first pioneering demonstrations of ultra thin light emitting diodes, tunnelling transistors, photodiodes, and solar cells, have recently been reported\cite{Withers2015,yu2013highly,britnell2012field,Britnell2013,massicotte2016picosecond}. In principle, the vdWH concept combined with the richness of the 2D materials compound space, makes it possible to design materials with atomic-scale control over the electronic states and energies. Leaving aside challenges related to the fabrication of such complex heterostructures, the basic design principles for such an approach remain to be established. In general, the electronic properties of a vdWH, such as the band structures of the constituent 2D crystals and band alignment across interfaces, depend  on many factors including interlayer hybridization, charge transfer, dielectric screening, proximity induced spin-orbit interactions, etc. Adding to this the complex and often uncontrolled atomic structure of the van der Waals interfaces, it becomes clear that quantitative prediction of energy levels in vdWHs is a delicate problem. In this work we take a first step in this direction by developing a method for calculating the change in the energy levels of a 2D material induced by the change in dielectric screening when it is embedded in a vdWH. 

The gold standard for quasiparticle (QP) band structure calculations of solids is the many-body GW method. It was introduced by Hedin in 1965\cite{hedin1965new} and first implemented and applied in an ab-initio framework two decades later\cite{hybertsen1985first, godby1986accurate}. Over the years, the GW approximation, in particular its non self-consistent G$_0$W$_0$ version, has been widely applied to bulk solids\cite{shishkin2006implementation} and molecules\cite{rostgaard2010fully, blase2011first}, and its superior performance over density functional theory (DFT) methods is well established. More recently, the \gowo method has been applied to 2D materials\cite{tran2014layer, rasmussen2015computational}, and technical issues related to $k$-point convergence and spurious interlayer screening arising from periodic boundary conditions have been addressed\cite{rasmussen2016efficient}. As an example, well converged G$_0$W$_0$ band gaps reported for monolayer MoS$_2$ are in the range 2.5 - 2.7 eV\cite{rasmussen2016efficient, qui2015erratum} in good agreement with the experimentally reported value of $\sim 2.5$ eV for the freestanding monolayer\cite{bolotin2014}. In comparison, DFT with the standard PBE functional yields a band gap of only 1.7 eV\cite{qiu2013optical}. Replacing the PBE by the HSE screened hybrid functional significantly improves the DFT band gap to 2.3 eV\cite{ellis2011indirect}. However, as explained below, the short-range nature of the DFT functionals, makes them incapable of describing
changes in QP band energies arising from screening of the electron-electron interaction by charges located outside the material, i.e. environmental screening. As shown in this work, the ultra thin nature of 2D materials combined with their poor intrinsic screening makes this effect particularly important in vdWHs.  

The effect of environmental screening on QP energy levels has been explored in great detail for molecules on surfaces. For example, GW calculations predict a 3 eV reduction in the HOMO-LUMO gap of benzene when physisorbed on graphite\cite{neaton2006renormalization}. Recalling that the QP energy levels represent electron removal and addition energies, the closing of the gap can be easily understood as a result of the charged final state of the molecule interacting with its image charge in the surface. The image charge effect has been confirmed in several experiments and is essential to include for a correct description of energy level alignment and charge transport at metal-molecule interfaces\cite{strange2011self}. 

A naive interpretation of the image charge effect would suggest that it becomes unimportant for a 2D material because the electrostatic interaction between the completely delocalised 2D charge distribution of a Bloch state and its equally delocalised image charge, scales as $1/A$ where $A$ is the area of the 2D material. However, this picture is incorrect because it is not the electron that is delocalised but its probability density. In fact, G$_0$W$_0$ calculations have found that the QP band gap of hBN is reduced by 1.1 eV when physisorbed on a graphite substrate.\cite{huser2013quasiparticle}. For monolayer MoSe$_2$, G$_0$W$_0$ predicts a somewhat smaller gap reduction of 0.15 eV for a similar substrate\cite{ugeda2014giant}. From these results it is already clear that the band gap renormalization in 2D materials is highly system dependent and cannot be described by a simple image charge model. That is, a simple image charge model cannot explain the large differences in the band gap reduction for benzene (3 eV), hBN (1 eV), and MoSe$_2$ (0.15 eV) when placed on a graphite substrate. In this work we show that these differences arise because the intrinsic screening in the three systems is very different and therefore the \emph{relative} effect of substrate screening becomes progressively smaller for benzene, hBN, and MoSe$_2$. 

The discussion above shows that the GW approximation must be considered the method of choice for quantitative and predictive descriptions of QP band structures in vdWHs. A practical problem, however, is that GW calculations have very high computational cost that grows quickly with system size. In the case of vdWHs, very large in-plane unit cells are typically required due to lattice mismatch between different 2D crystals. This implies that GW calculations for simple bilayer structures become unfeasible unless the layers are strained or compressed to fit in a small common unit cell. Since 2D materials band structures are high sensitive to strain\cite{yun2012thickness} this solution can, however, only be used for qualitative studies.

An alternative strategy that we follow here is to neglect the electronic hybridization between the layers in the GW calculation thereby circumventing the need for large supercells. Interlayer hybridization is generally weak in vdWHs. This is true in particular for (nearly) incommensurate interfaces where the effect of hybridisation on the band structure has been found to be negligible.\cite{Woods2014, lu2014mos} In cases where hybridisation is important, i.e. for commensurate interfaces, it can be treated separately using lower level methods such as DFT or tight-binding. Indeed, using the case of multilayer MoS$_2$ as an illustration, we show here that this approach works surprisingly well. To account for the effect of environmental screening on the QP band structure of a given layer, we use the Quantum Electrostatic Heterostructure (QEH) model\cite{andersen2015dielectric}, to compute the change in the screened Coulomb interaction ($\Delta W$) within the layer due the surrounding layers of the heterostructure. This allows us to evaluate the change in the GW self-energy and from this compute the corrections to the QP energies of the layer. In the following we give a more detailed presentation of this method.  


\begin{figure*}[!]
  \includegraphics[width = 0.8\linewidth]{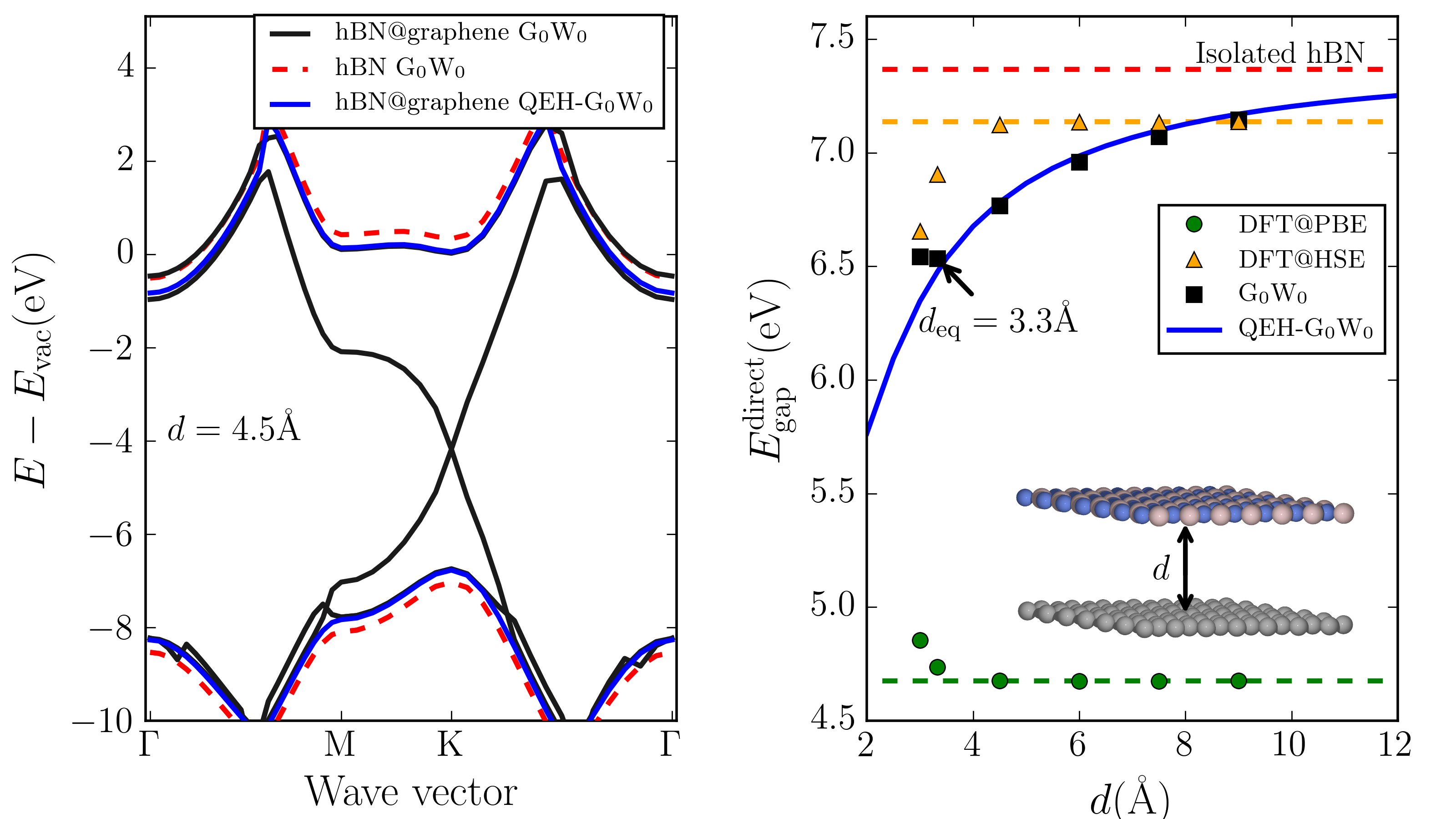}
   \caption{Top: Band structures of bilayer hBN@graphene at separation $d=4.5$\AA, calculated with \gowo (grey dashed) and G$_0$$\Delta$W$_0$ (blue), together with the \gowo band structure for freestanding hBN (red dashed). Excellent agreement is found with G$_0$$\Delta$W$_0$ perfectly reproducing the reduction in band gap due to graphene. Bottom: hBN band gap at $\mathrm{K}$ as a function of hBN-graphene binding distance, $d$, calculated with PBE (black), HSE (blue) and \gowo(red). The horizontal dashed lines indicate band gap of freestanding hBN. The G$_0$$\Delta$W$_0$ result (red solid line) is again in excellent agreement with the \gowo result (except for the smallest $d$ which is below the equilibrium separation) and captures the long-ranged image-charge effect on the band gap. Clearly, this is not the case for the DFT methods, where the band gap is constant for $d > 4 \AA$.  
}
 \label{fig1}
\end{figure*}

\begin{figure*}[t]
  \includegraphics[width = 0.9 \linewidth]{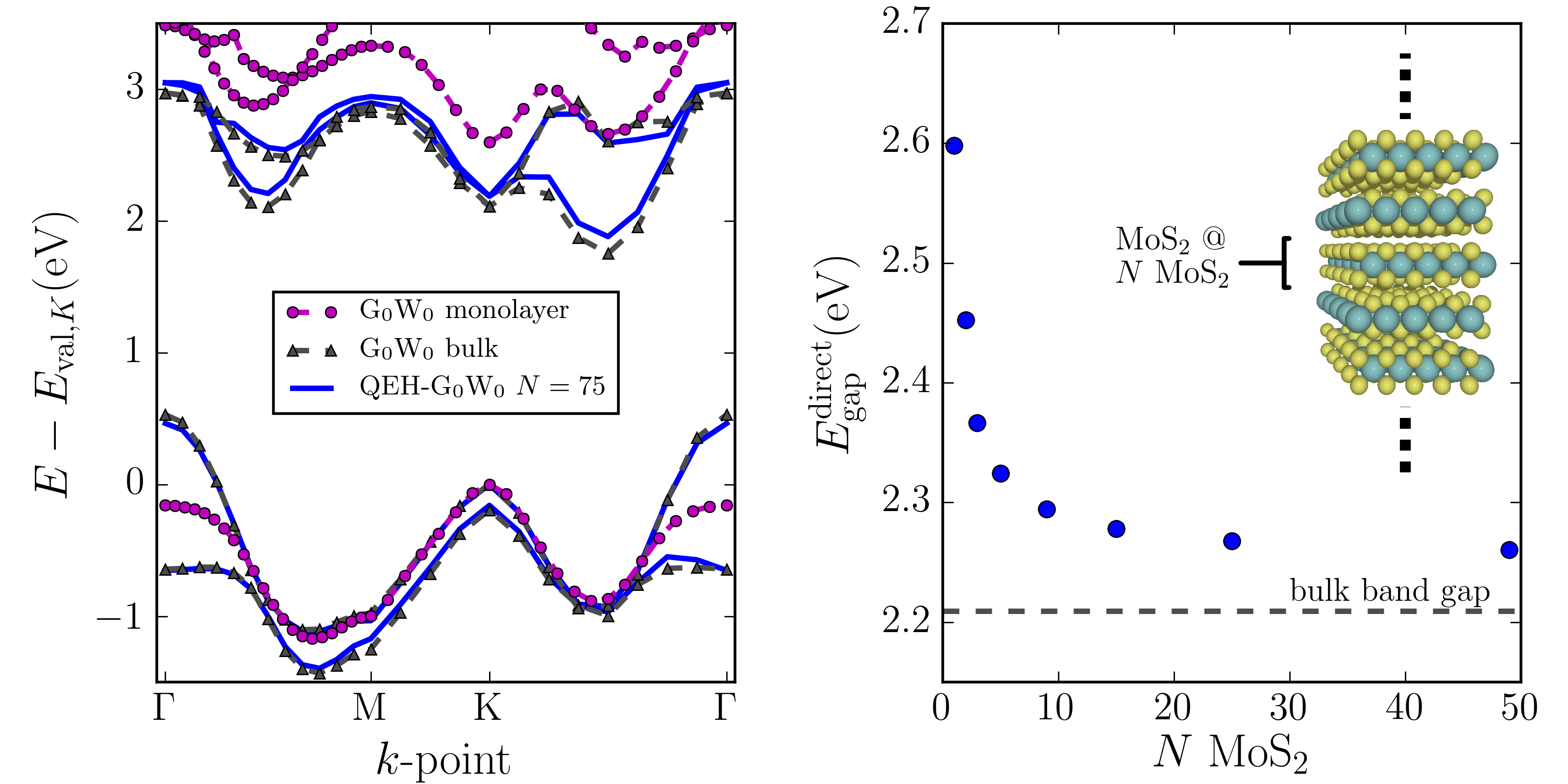}
   \caption{Left: \gowo band structures of monolayer (purple dots) and bulk (black triangles) MoS$_2$, compared to the G$_0$$\Delta$W$_0$ result calculated for a layer of MoS$_2$ embedded in the middle of a 75 layer thick MoS$_2$ slab. Right: Band gap at the $K$-point of an MoS$_2$ layer embedded in progressively thicker MoS$_2$ slabs.}
 \label{fig3}
\end{figure*}

The basic principle of the G$\Delta$W method is to calculate the \textit{change} in QP band energy of a particular layer due the change in the (long-ranged) electronic screening originating from the surrounding layers. This can be formally expressed in terms of the screened Coulomb interaction,
\begin{equation}\label{eq:epsilon}
W(\rr, \rr', \omega) = \int d\rr'' \epsilon^{-1}(\rr,\rr'', \omega) V(\rr'', \rr'),
\end{equation}
where $V(\rr, \rr' ) = \frac{1}{|\rr - \rr'|}$ is the bare Coulomb potential and $\epsilon(\rr,\rr', \omega)$ is the microscopic dielectric function that can be obtained from first principles\cite{Yan2011a}. The QEH model evaluates the dielectric function and screened potential of a vdWH following a two-step procedure: First the dielectric building blocks (a low dimensional representation of the $\mathbf q_{\parallel}$-dependent dielectric function) of the isolated layers forming the heterostructure are calculated from first principles or simply imported from a database. Next, the dielectric function of the entire heterostructure is obtain by solving a simple set of linear equations describing the electrostatic coupling between the layers. We refer to Ref.~\cite{andersen2015dielectric} for details on the QEH method.

The QP energy of an electronic state, $\ket{n \mb{k}}$, belonging to layer $i$ of a heterostructure (HS), can be expressed as
\begin{equation} \label{eq:dEquasiQEH}
E^{i, \text{HS}}_{n \mb{k}} = E^i_{n \mb{k}} (\m{G}_0\m{W}_0) + \Delta E^{i, \text{scr}}_{n \mb{k}}+ \Delta E^{i, \text{hyb}}_{n \mb{k}} 
\end{equation}
where $E^i_{n \mb{k}} (\m{G}_0\m{W}_0)$ is the QP energy of the freestanding monolayer which can be calculated once and for all, and $\Delta E^{i, \text{scr}}_{n \mb{k}}$ and $\Delta E^{i, \text{hyb}}_{n \mb{k}}$ are the corrections due to environmental screening and interlayer hybridisation, respectively. The present work is concerned with the calculation of the screening term, although the role of the hybridisation term will also be addressed. 

In the standard \gowo method, the QP energies are obtained on top of DFT eigenvalues as
\begin{equation} \label{eq:Equasi}
E_{n \mb{k}} = \epsilon_{n \mb{k}} + Z_{n \mb{k}} \cdot \m{Re} \, \bigl[\Sigma_{n \mb{k}}(\omega =  \epsilon_{n \mb{k}} ) + V_{n \mb{k}}^{\m{EXX}}- V_{n \mb{k}}^{\m{XC}}\bigr],
\end{equation}
where $\epsilon_{n \mb{k}}$ is the DFT eigenvalue, $V_{n \mb{k}}^{\m{EXX}}$ the exact exchange contribution, $V_{n \mb{k}}^{\m{XC}}$ the exchange-correlation potential, $\Sigma_{n \mb{k}}$ is the GW self-energy, and $Z_{n \mb{k}}$ the so-called renormalization factor. In the G$_0$W$_0$ method, the self-energy takes the form  
\begin{equation}\label{eq:sigma}
\Sigma(\rr, \rr', \omega) = \frac{i}{2\pi}\int d\omega' G_0(\rr, \rr', \omega + \omega') \overline{W}_0(\rr, \rr', \omega'),
\end{equation}
where $G_0$ the single-particle GreenÕs function of an approximate DFT Hamiltonian and $\overline{W}_0 = W_0 - V $ is the difference between the screened and the bare Coulomb interaction (the exchange part coming from $V$ is treated separately via the $V_{n \mb{k}}^{\m{EXX}}$ term). The screened interaction is calculated within the random phase approximation using the non-interacting $G_0$.

In order to treat the screening from the surrounding medium separately, we split the screened potential into two parts (dropping the 0-subscript on $W$ for notational simplicity)
\begin{equation}\label{eq:Wsplit}
\overline{W}_{i,\m{ HS}}(\rr, \rr', \omega) = \overline{W}_{i}(\rr, \rr', \omega) + \Delta \overline{W}_{i}(\rr, \rr', \omega).
\end{equation}
where $\overline{W}_{i}$ is the screened potential of the freestanding monolayer and $\Delta \overline{W}_{i}$ is the change in screening due to the surrounding layers.
Inserting the first term into Eqs.~(\ref{eq:sigma}) and (\ref{eq:Equasi}) gives the standard expression for the \gowo eigenvalue of the freestanding monolayer under the assumption that the change in the renormalization factor can be neglected, $Z^i_{n \mb{k}} \approx Z^{i, HS}_{n \mb{k}}$. The second term of Eq.~(\ref{eq:Wsplit}) is evaluated by the QEH and used to get the resulting change in the QP energy of an electronic state $\ket{n \mb{k}}$ belonging to layer $i$:  
\begin{align} \label{eq:dEquasi}
&\Delta E^{i, \text{scr}}_{n \mb{k}} = Z^{i}_{n \mb{k}}  \cdot \m{Re} \, \Delta  \Sigma^{i}_{n \mb{k}} (\omega=  \epsilon_{n \mb{k}} ), 
\end{align}
where the change in self-energy, $\Delta  \Sigma$, is obtained from Eq.~(\ref{eq:sigma}), by inserting the change in screened potential, $\Delta \overline{W}$. More details are provided in the Methods section. 
 

Physically, $\Delta  \Sigma^{i}$ represents the potential felt by an electron in layer $i$ due to the polarisation of the surrounding layers caused by the electron itself. Since the spatial variation of this potential within the layer is relatively weak, the effect of $\Delta  \Sigma^{i, \text{HS}}$ is very similar for all states and it mainly causes a $\mb{k}$-independent shift of the energy bands which is positive for occupied states and negative for unoccupied states, i.e. a symmetric gap reduction.  

We note that previous methods, similar in spirit to the one proposed here, have used model dielectric functions to approximate the difference between the GW self-energy and the LDA potential\cite{gygi1989quasiparticle,rohlfing2010electronic}. These methods are more general than the present, which is explicitly designed for vdWHs. On the other hand, for this specific class of materials, our method should be superior due to the use of ab-initio rather than model dielectric functions for calculating $\Delta W$.   

In the following we present calculations of band structures and band gaps calculated with the G$_0$$\Delta$W$_0$ approach, and compare to results obtained from standard \gowo calculations in order to evaluate the performance of the method. Due to the large computational cost of the \gowo approach we restrict these calculations to layer combinations where it is reasonable to to construct lattice matched structures in the simple unit cell. First, we consider the case of monolayer hBN on top of a single layer of graphene, see Fig.~\ref{fig1}.

 The calculated band structures of monolayer hBN and the hBN@graphene bilayer at separation $d=4.5$\AA are shown in Fig.~\ref{fig1}. All energies are aligned relative to vacuum. In all these calculations only the screening contribution to the QP energy has been included in Eq. (\ref{eq:dEquasiQEH}). The G$_0$$\Delta$W$_0$ band structure (blue) is seen to be in excellent agreement with the full \gowo calculation (black). Comparing to the band structure of the isolated hBN layer (red dashed) it is clear that the effect of the graphene layer is an (almost) constant and symmetric closing of the hBN band gap of 0.6 eV. This example demonstrates that the G$_0$$\Delta$W$_0$ method is essentially exact in the weak-hybridisation limit. For smaller layer distances, the effect of hybridzation leads to an opening of the hBN gap at the $\Gamma$ point while the bands in the remaining part of the BZ are almost unaffected down to the equilibrium layer distance of $3.3$\AA. We note that the effect of hybridization is expected to be particularly large for lattice matched interfaces as the present. 
Several studies of 2D bilayers have shown that the degree of hybridization between the layers depends on the relative rotation \cite{heo2015interlayer, constantinescu2015energy, wang2015electronic}, where hybridization has been found to diminish as the layers are rotated away from perfect matching. Thus, we expect that for incommensurate or non-lattice matched systems, the degree of hybridization will be less pronounced, and the G$_0$$\Delta$W$_0$ method will be an even better approximation. Nevertherless, as will be shown later in this work, the effect of hybridisation can be treated separately at the DFT level and added to the G$_0$$\Delta$W$_0$ band structure.

In the right panel of Fig.~\ref{fig1} we show the direct band gap of hBN at the $\m{K}$-point as a function of binding distance to graphene. As expected, the band gap increases with increasing separation and converges towards the band gap of freestanding hBN. The distance dependence closely follows a $1/d$ behaviour reflecting the image charge nature of the phenomenon. There is an excellent agreement between the \gowo and G$_0$$\Delta$W$_0$ results for all distances down to the equilibrium distance of $3.3$\AA. 

For comparison we show the DFT single-particle band gap of hBN obtained with the PBE (blue) and HSE (blue) xc-functionals. As expected, the HSE yields band gaps in much better agreement with the G$_0$W$_0$ result than the PBE. However, it is also clear that neither of the DFT functionals capture the effect of screening from the graphene sheet and the band gap does not change for distances beyond 4.0 \AA.   

 

Having verified the accuracy of the G$_0$$\Delta$W$_0$ method for weakly hybridising layers we now turn to the case where the effects of hybridisation and screening are of similar magnitude. As an example, we consider the change in the band structure of monolayer MoS$_2$ when inserted into a thick slab of MoS$_2$, i.e. the 2D-3D transition of the MoS$_2$ band structure. It is well known that monolayer MoS$_2$ is a direct band gap semiconductor with a QP band gap at the $K$-point of 2.5 eV\cite{rasmussen2016efficient} while the bulk has an indirect gap of 1.3 eV. The qualitative change from direct to indirect band gap has been ascribed to interlayer hybridisation (also referred to as quantum confinement) that shifts the bulk valence band maximum at $\Gamma$ upwards in energy. However, a more quantitative assessment of the changes, including an analysis of the relative importance of hybridisation and screening, has not been reported before.

\begin{figure}
  \includegraphics[width =0.9 \linewidth]{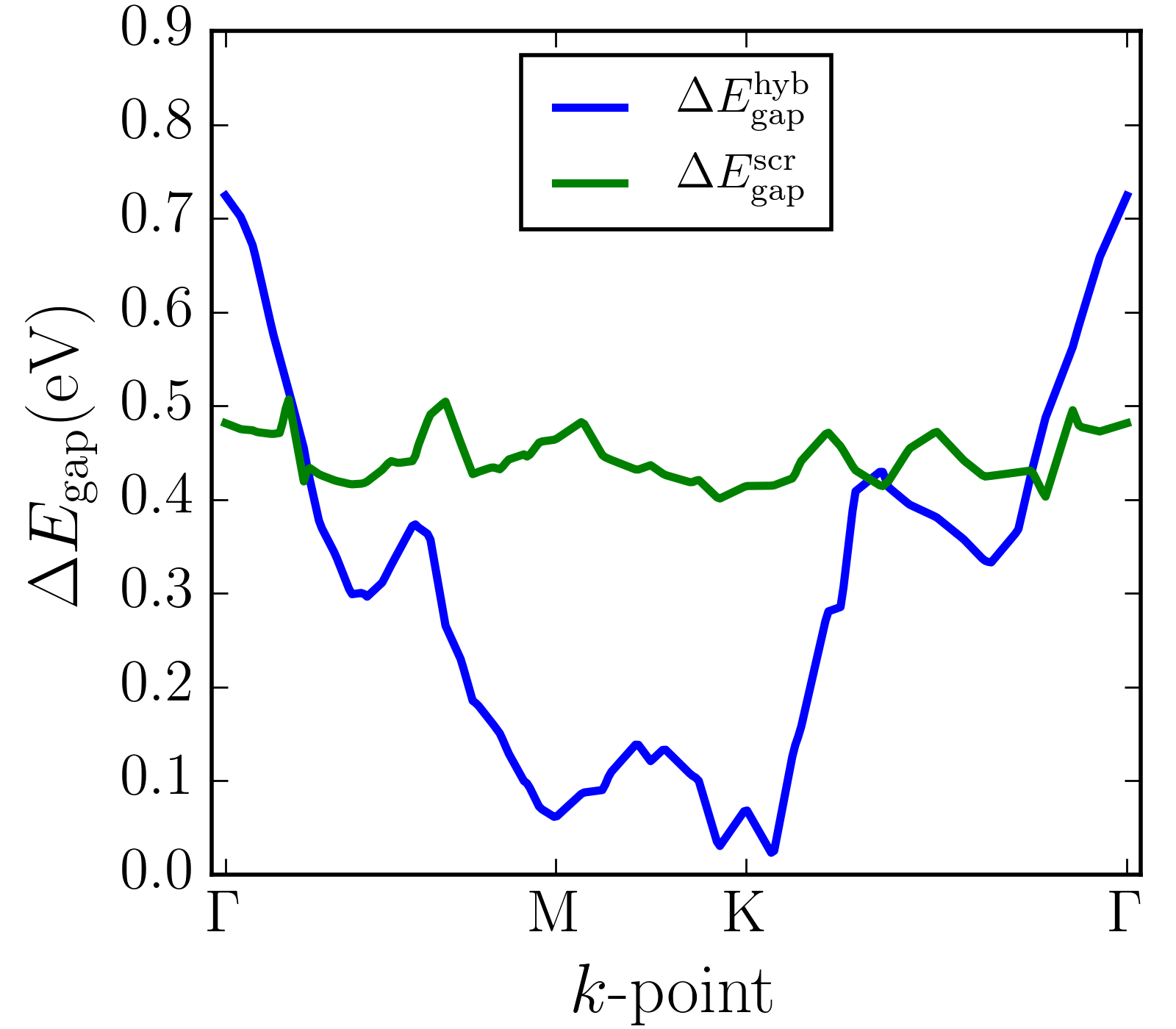}
   \caption{The change in band gap between monolayer MoS$_2$ and bulk MoS$_2$ has contributions from dielectric screening (green) and interlayer hybridisation (blue). The screening contribution is relatively constant throughout the BZ as the change in potential ($\Delta W$) due to the surrounding layers, has weak spatial variation. In contrast the hybridisation shows strong $k$-dependence reflecting the change in orbital character of the wave functions from Mo-$d$ ($K$) to S-$p$ ($\Gamma$).}
 \label{fig5}
\end{figure}

The effect of interlayer hybridisation is simply extracted from a DFT calculation, i.e. 
\begin{equation}
\Delta E_{n\mb{k}}^{\text{hyb}} =  E_{n \mb{k}}(\m{DFT, bulk})  - E_{n \mb{k}}(\m{DFT, monolayer}), 
\end{equation}
where we align the valence bands at the $K$-point. It should be noted that this approach assumes that the effect of screening is negligible in the DFT calculation so that this effect is not double counted when $\Delta E_{n\mb{k}}^{\text{scr}}$ is added. This was indeed the conclusion of the previous analysis of the hBN/graphene bilayer see Fig. \ref{fig1}. 
It should also be noted that we take the $\mb{k}$-vector to lie in the two-dimensional BZ of the monolayer, because the out-of-plane momentum is not a good quantum number for general vdWHs. 

\begin{figure}
  \includegraphics[width = 0.9 \linewidth]{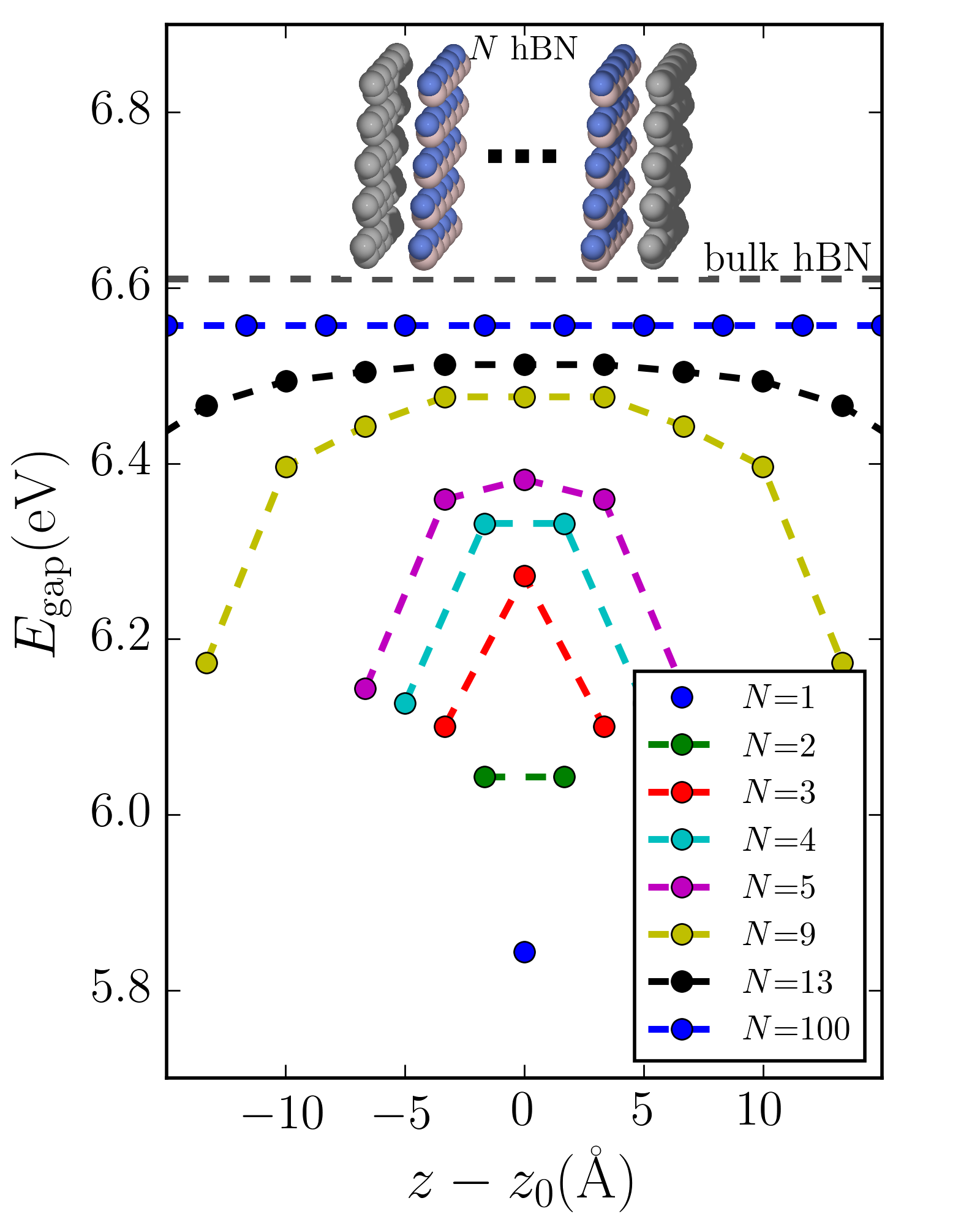}
   \caption{The layer-specific band gap evaluated at different positions of an $N$-layer hBN stack sandwiched between two graphene sheets. $z_0$ denotes the center of the structure and the markers represent the position of the hBN layers. The screening correction is seen to be larger for the hBN layers close to graphene. As the width of the structure increases, the band gap at the center of the slab converges to that of bulk hBN (dashed horizontal line).}
 \label{fig2}
\end{figure}

Fig.~\ref{fig3}(left) shows the band structures of monolayer and bulk MoS$2$ calculated with the G$_0$W$_0$ (purple and black lines, respectively). Apart from small deviations of less than $0.1$ eV, the G$_0$$\Delta$W$_0$ band structure, evaluated for a monolayer in the center of a 75 layer thick MoS$_2$ slab, is an excellent approximation to the G$_0$W$_0$ result for the bulk. The right panel shows the evolution of the G$_0$$\Delta$W$_0$ calculated direct band gap at the $K$-point of a monolayer MoS$_2$ as a function of the number of surrounding MoS$_2$ layers. The screening leads to a gap reduction of around 0.35 eV for a 50 layer slab, in good agreement with the 0.4 eV reduction predicted by comparing the G$_0$W$_0$ band gaps for the bulk and monolayers at the $K$-point. From this analysis we draw two conclusions: (i) The effects of interlayer screening and hybridisation in vdWHs can, to a very good approximation, be considered as independent and be treated separately. (ii) The difference between the band structure of monolayer and bulk MoS$_2$ is composed of a screening induced $\mathbf k$-point independent, symmetric band gap closing of $\sim0.4$ eV, and a highly $\mathbf k$-point dependent shift due to hybridisation, see Fig. \ref{fig5}.   

\begin{figure*}
  \includegraphics[width = 0.9 \linewidth]{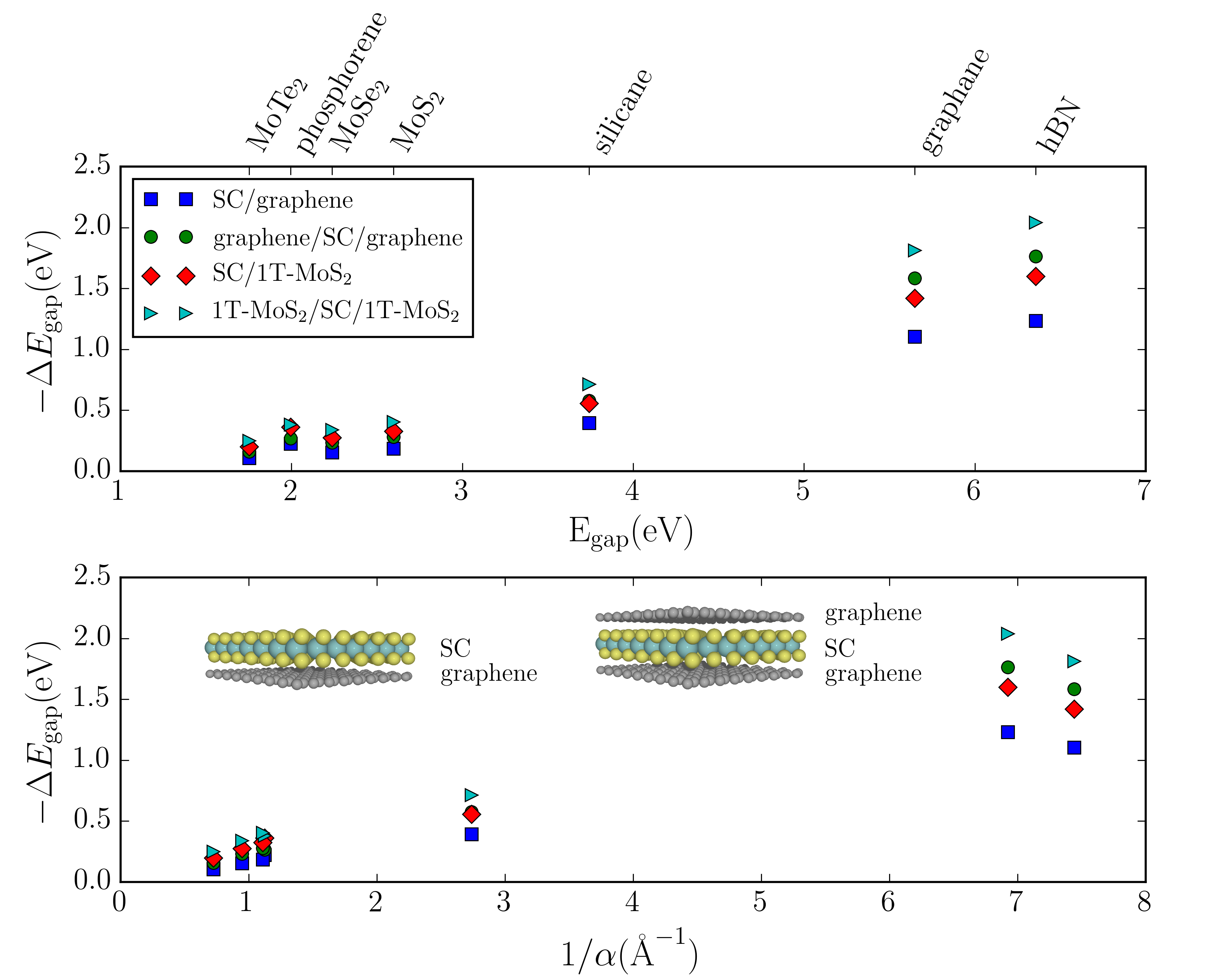}
   \caption{G$_0$$\Delta$W$_0$ calculated band gap correction ($\Delta E_\m{gap}$) for seven 2D semiconductors (SC) in four different heterostructure configurations, namely on top of a single graphene sheet (blue), sandwiched between two graphene sheets (green) and the same configurations with graphene replaced by the metallic 1T phase of MoS$_2$. The two panels shows the band gap reduction plotted versus the freestanding 2D semiconductor band gap and inverse static polarizability, respectively. }
 \label{fig4}
\end{figure*}

Having assessed and validated the accuracy of the G$_0$$\Delta$W$_0$ method, we next apply it to investigate band structures in vdWHs that are beyond reach with standard \gowo calculations. As a first example we have considered the variation of the band gap in a heterostructure consisting of an hBN multilayer film sandwiched between two graphene sheets. Fig.~\ref{fig2} shows the band gaps of the individual hBN layers as a function of the position of the layers from the centre of the film. As expected, the screening correction to the gap is larger in the vicinity of graphene, resulting in a smaller band gap towards the end of the hBN film. As the thickness of the structure increases, the effect of graphene vanishes in the middle of the structure, and the band gap converges towards that of bulk hBN. This example demonstrates how dielectric screening can be exploited to tune the band gap of 2D materials \emph{via} van der Waals heterostructuring.   

Finally we employ the G$_0$$\Delta$W$_0$ method to address the question of the strong system dependence of the screening induced band gap renormalization. As mentioned in the introduction, the reduction in the energy gap of a benzene molecule, an hBN monolayer, and an MoS$_2$ monolayer, when placed on a graphite surface has been reported to be 3 eV, 1.1 eV and 0.15 eV, respectively. This shows that the reduction cannot be explained by a model that only considers the interaction between the bare electron/hole with its image charge in the substrate, since this would yield the same reduction for all three systems. 

We have used the G$_0$$\Delta$W$_0$ method to calculate the band gap reduction of seven different 2D semiconductors (SC) in four different heterostructure configurations, namely on top of a graphene layer, sandwiched between two graphene layers, and the same configurations with graphene replaced by the metallic 1T phase of MoS$_2$, see Fig.~\ref{fig4}. In the top panel the band gap correction is plotted relative to the absolute \gowo band gap of the freestanding 2D semiconductors. Generally, the band gap correction is larger for the 1T-MoS$_2$ substrate than for graphene reflecting the increased screening from a metallic substrate compared to the semi-metallic graphene. The magnitude of the band gap correction is seen to scale with the size of the 2D SCÕs band gap: Thus for large band gap materials, such as hBN and graphane, the correction is in the 1-2 eV range depending on the substrate/embedding while for medium band gap semiconductor such as MoS$_2$, the band gap correction is in the range 0.2 - 0.4 eV.

In the lower panel of Fig.~\ref{fig4}, the band gap correction is plotted against the inverse (static) in-plane polarisability, $1/\alpha$. Again a very direct scaling is observed. This is not surprising as the polarisability of a 2D semiconductor has been found to scale roughy inversely with its band gap\cite{olsen2016simple}. For small in-plane wave vectors, $q$, the polarizability of an isotropic 2D semiconductor is related to the dielectric function by 
\begin{equation} \label{eq:alpha}
\epsilon(q,\omega) = 1 + q \alpha(\omega). 
\end{equation}
Note that the definition and first-principles calculation of the dielectric function of a (quasi-) 2D material is different from the usual case of bulk materials. In particular, the total potential should be averaged only over the physical width of the 2D material rather than the entire supercell which would yield $\epsilon(q,\omega)\to 1$ as the supercell height is increased\cite{Huser2013a}. A larger (smaller) value of $\alpha$ corresponds to stronger (weaker) \emph{intrinsic} screening in the 2D semiconductor. It is then clear that the reason for the smaller band gap reduction observed in 2D materials with larger polarisability, or smaller native band gap, is the smaller \emph{relative} reduction in the screened interaction caused by the environment. In small molecules, the intrinsic screening is vanishingly small and classical image charge models capture well the reduction in the energy gap\cite{garcia2009polarization}. For 2D semiconductors, the intrinsic screening is of intermediate strength and environmental screening has a smaller yet sizeable and less trivial effect on the band structure.     

The QEH and G$\Delta$W codes are freely available as part of the GPAW open source code. The QEH code calculates the dielectric building blocks that are used by the G$\Delta$W code to compute the change in screened potential, $\Delta \overline{W}_{i}$, and the corresponding QP energy corrections. Pre-calculated building blocks are available for a number of 2D materials as are the scripts for reproducing the results in this work.

The quantitatively accurate prediction of quasiparticle band structures of vdWHs is a very complex challenge. The G$\Delta$W method presented here focuses on the effect of environmental screening. However, a full description should also account for a number of other effects including interlayer hybridization and charge transfer, doping, and proximity induced spin-orbit coupling. The assumption underlying this work is that these effects are relatively independent and therefore can be treated separately employing the best suited methods. More work is required to test this assumption and to further improve and develop the methodology. Obviously, access to reliable experimental benchmark data will play an essential role for this development. 

In conclusion, we have presented a method for calculating quasiparticle band structures of 2D materials embedded in van der Waals heterostructures. The method combines the quantum electrostatic heterostructure (QEH) model for the dielectric properties of vdWHs with the many-body GW method for QP band structures. Our approach significantly reduces the computational cost of standard GW schemes and overcomes the practical problems encountered when applying such schemes to lattice mismatched heterostructures. 
Using the G$\Delta$W method we performed a systematic investigation of the role of environmental screening on the band gaps of different 2D semiconductors and showed how it can be utilised to control the QP band structure of 2D materials \emph{via} vdWH design. 

\subsection{Methods}
In plane wave representation, the GW self-energy is expressed as: 
\begin{widetext}
\begin{align} \label{eq:selfgpaw}
 & \Sigma_{n \mathbf{k}} \equiv \left<n \mathbf{k} \middle| \Sigma(\omega = \epsilon_{n \mathbf{k}}) \middle|n \mathbf{k} \right>    = \nonumber \\
& \frac{1}{(2\pi)^3}\int_\text{BZ} d\mathbf{q} \sum_{\mathbf{GG}'} \frac{i}{2\pi} \int_{-\infty}^\infty d\omega' \overline{W}_{\mathbf{GG}'}(\mathbf{q}, \omega') 
\times \sum_m \frac{[\rho_{n,\mathbf{k}}^{m,\mathbf{k} + \mathbf{q}}(\mathbf{G})] [\rho_{n\mathbf{k}}^{m,\mathbf{k}+\mathbf{q}}(\mathbf{G}')]^*}{\omega + \omega' - \epsilon_{m\mathbf{k} + \mathbf{q}} - i\eta\,\text{sgn}(\epsilon_{m\mathbf{k} + \mathbf{q}}-\mu)},
\end{align}
\end{widetext}
where $\mu$ is the chemical potential, $\eta$ is a broadening parameter, and $\rho_{n,\mathbf{k}}^{m,\mathbf{k} + \mathbf{q}}(\mathbf{G})  = 
\left<n \mathbf{k} \middle| e^{i(\mathbf{q} +
\mathbf{G})\mathbf{r}} \middle|m \, \mathbf{k} \!-\! \mathbf{q} \right>$ is the pair density matrix. The screened potential (precisely: the screened potential minus the bare potential), $\overline{W}_{\mathbf{GG}'}(\mathbf{q}, \omega) $, is calculated from 
\begin{equation}
\overline{W}_{\mathbf{G} \mathbf{G}'}(\mathbf{q}, \omega) = (\epsilon^{-1}_{\mathbf{G} \mathbf{G}'}(\mathbf{q}, \omega) - \delta_{\mathbf{G} \mathbf{G}'}) V_{\mb{G}}(\mathbf{q}),
\end{equation}
where $V_{\mb{G'}}(\mathbf{q})$ is the bare Coulomb potential and the dielectric matrix, $\epsilon_{\mathbf{G} \mathbf{G}Õ}$, is calculated within the random phase approximation (RPA).


In the G$\Delta$W method the macroscopic component of the change in screened potential, $\Delta \overline{W}(\q, \omega)$, due to the surrounding layers is calculated using the QEH method. Specifically, the difference in screened potential of the freestanding monolayer, $i$, and the embedded layer is expressed
\begin{align} \label{eq:dW}
&\Delta \overline{W}_i(\q, \omega) = \nonumber \\ 
& \frac{1}{d_i^2} \iint \displaylimits_{z_i-d_i / 2}^{z_i+d_i/2} \overline{W}_{\mathrm{HS}}(z, z', \q, \omega) -  \overline{W}_i(z, z', \q, \omega) dz dz',
\end{align}
where $z_i$ and $d_i$ are the centre and width of the layer, respectively. Since $\Delta \overline{W}(z, z', \q, \omega)$ turns out to be a slowly varying function of $z$ and $z'$ (because it is generated by charge distributions located outside layer $i$), it is generally a good approximation only to include only the constant (monopole) component of the screening. With this approximation, the change in the GW self-energy for a state in layer $i$, becomes
\begin{widetext}
\begin{align*} 
\Delta \Sigma_{n \mathbf{k}}^i   \approx 
 \frac{1}{(2\pi)^3}\int_\text{BZ} d\mathbf{q}  \frac{i}{2\pi} \int_{-\infty}^\infty d\omega' \Delta \overline{W}_i(\mathbf{q}, \omega')  \sum_m \frac{[\rho_{n,\mathbf{k}}^{m,\mathbf{k} + \mathbf{q}}(\mathbf{G=0})] [\rho_{n\mathbf{k}}^{m,\mathbf{k}+\mathbf{q}}(\mathbf{G'=0})]^*}{\omega + \omega' - \epsilon_{m\mathbf{k} + \mathbf{q}} - i\eta\,\text{sgn}(\epsilon_{m\mathbf{k} + \mathbf{q}}-\mu)},
\end{align*}
\end{widetext}
where only the macroscopic $\mathbf{G=G'=0}$ component of the pair density matrices are required. 

\subsection{Computational details.}
All calculations were performed using the GPAW electronic structure code\cite{GPAW2} which implements linear response theory and the GW approximation in a plane wave basis\cite{Yan2011a, huser2013quasiparticle}. All DFT calculations for the 2D materials were performed with a plane wave cutoff of 600 eV and a $k$-point grid of $42\times 42$, except for phosphorene where $30 \times 42$ $k$-points were used. At least 15 \AA of vacuum was added to separate the layers in the perpendicular direction. The G$_0$W$_0$@PBE calculations for 2D materials used the same $k$-point grid as the DFT calculations. 

The self-energy and screened interaction was calculated for reciprocal lattice vectors up to $200$ eV which was also used as cut-off for the sum over empty states. Results were extrapolated to the infinite plane wave limit as described in Ref. \cite{klimevs2014predictive}. A truncated Coulomb interaction was employed to avoid artificial screening due to the periodically repeated layers\cite{Rozzi2006}. The G$_0$W$_0$@PBE calculations for bulk hBN and 2H-MoS$_2$ employed $24\times 24\times 8$ and $24\times 24\times 6$ $k$-points, respectively.    

The QEH dielectric building blocks of graphene and 1T-MoS$_2$ were calculated on dense $k$-point grids of $200\times 200$ and $160\times 160$, respectively, to obtain a dense sampling of the screened potential in the $q \rightarrow 0 $ limit. The dielectric building blocks for the semiconducting materials were calculated on $42\times 42$ $k$-point grids and interpolated to the finer $k$-point grids employing an analytical expansion of the response function valid for gapped systems around $q=0$.    

In-plane lattice constants were obtained by structural relaxations using the PBE functional. Out of plane binding distances between layers were estimated based on available experimental and computational data for the layered bulk crystals. Specifically, the layer separation for an A-B heterostructure was obtained as the average layer-layer distance of the A and B bulk crystals. This approach was taken to avoid the complex problem of computing interlayer distances for general lattice mismatched van der Waals heterostructures. All the structural parameters used in this work are listed in Table \ref{table:lattice}. 

\section*{Acknowledgement}
The author gratefully acknowledge the financial support from the Center for Nanostructured Graphene (Project No. DNRF103) financed by the Danish National Research Foundation.

The authors declare no competing financial interest.
\begin{table*}[]
\small
\centering
\begin{tabular}{l |c |c |c |c|c|c|c|c|c}
Material & graphene & 1T-MoS$_2$ & 2H-MoTe$_2$ & 2H-MoSe$_2$ & 2H-MoS$_2$ & phosphorene & silicane & graphane & hBN \\
\hline
In-plane lattice constant (\AA) & 2.46  & 3.193 & 3.547 & 3.320 &  3.184 & 4.630 / 3.306 & 3.892  & 2.541 & 2.54\\
Out of plane binding distance (\AA)&  3.354 & 6.332  & 7.220 & 6.676 &  6.253 & 6.400 &  6.923 &  4.978 & 3.326\\
Binding distance to graphene (\AA)& --  & -- & 5.287 & 5.015 & 4.803 & 4.877 &  5.138 &  4.1659 & 3.336 \\
Binding distance to 1T-MoS$_2$ (\AA)& -- & -- & 6.776  & 6.504 & 6.293 & 6.366 &  6.627 & 5.655 & 4.829\\
\end{tabular}
\caption{Lattice constants and interlayer binding distances used in this study. For a consistent treatment of the metal dichalcogenides, the out of plane lattice constants are approximated as $c = 2 s_0$, where $s_0$ is the vertical distance between the chalcogenide atoms (S, Se, Te) of the monolayer. The bonding distance to graphene and 1T-MoS$_2$ is approximated as the average layer separation of the two (bulk) species.}
\label{table:lattice}
\end{table*}
 \FloatBarrier

\bibliography{bibtex}

\end{document}